\documentclass[12pt,draftcls,english,onecolumn]{IEEEtran}
\usepackage[T1]{fontenc}
\usepackage[latin9]{inputenc}
\usepackage{amsmath}
\usepackage{amssymb}
\usepackage{graphicx}
\usepackage{subfigure}

\makeatletter
\newtheorem{definitn}{Definition}
\newtheorem{prop}{Proposition}
\newtheorem{lemma}{Lemma}
\newtheorem{thm}{Theorem}

\usepackage{amsfonts}

\usepackage{array}

\usepackage{subfigure}

\usepackage{cite}

\usepackage{hyperref}

\newcommand{\set}[1]{\mathcal{#1}}
\newcommand{\Real}{{\mathbb{R}}}

\def\vecspace{{W}}
\def\field{{\mathbb F}}
\newcommand{\bb}[1]{{#1}}

\def\defined{{\: \triangleq \:}}

\def\setalpha{{\cal A}}

\def\N{{\cal N}}

\DeclareMathOperator{\con}{con}
\def\barcon{{\overline{\con}}}
\def\la{\langle}
\def\ra{\rangle}
\makeatother

\usepackage{babel}

\newcommand{\nequal}[1]{\stackrel{#1}{=}}
\newcommand{\myspan}[1]{\left\langle #1 \right\rangle}

\newcommand{\seq}[3]{#1_{#2},\ldots,#1_{#3}}

\newcommand{\X}{{\mathcal X}}
\newcommand{\V}{{\set{V}}}

\newcommand{\A}{\set{A}}
\newcommand{\B}{\set{B}}
\newcommand{\C}{\set{C}}

\newcommand{\reals}{\mathbb{R}}
\newcommand{\integers}{\mathbb{Z}}

\def\bff{{\bf f}}
\def\bfh{{\bf h}}
\def\bfg{{\bf g}}

\begin{document}

\title{Existence of new inequalities for representable polymatroids}

\author{Terence Chan, Alex Grant and Doris Kern}

\maketitle
\begin{abstract}
  An Ingletonian polymatroid satisfies, in addition to the polymatroid
  axioms, the inequalities of Ingleton (Combin. Math. Appln.,
  1971). These inequalities are required for a polymatroid to be
  representable. It is has been an open question as to whether these
  inequalities are also sufficient.  Representable polymatroids are of
  interest in their own right. They also have a strong connection to
  network coding. In particular, the problem of finding the linear
  network coding capacity region is equivalent to the characterization
  of all representable, entropic polymatroids.  In this paper, we
  describe a new approach to adhere two polymatroids together to
  produce a new polymatroid. Using this approach, we can construct a
  polymatroid that is not inside the minimal closed and convex cone
  containing all representable polymatroids. This polymatroid is
  proved to satisfy not only the Ingleton inequalities, but also the
  recently reported inequalities of Dougherty, Freiling and Zeger. A
  direct consequence is that these inequalities are not sufficient to
  characterize representable polymatroids.
\end{abstract}

\section{Introduction}
The idea of network coding was first proposed in the seminal
paper~\cite{Ahlswede.Cai.ea00network} as a means to increase
achievable transmission throughput in data communications networks. In
the traditional packet-switched routing approach, intermediate network
nodes can only duplicate received packets and forward them to
subsequent nodes.  In contrast, network coding allows arbitrary
computational data processing at intermediate nodes. For example,
intermediate nodes may forward arbitrary linear combinations of
several received packets. In the single source multicast scenario,
network coding significantly increases maximal transmission
throughput, and achieves the max-flow min-cut bound. It was
subsequently proved~\cite{Li.Yeung.ea03linear} that linear network
codes suffice to achieve maximal throughput for this case.  

While the easily computable maximum flow (and associated minimum cut)
determines the maximal attainable throughput in the single source
scenario, this bound is not tight in general (multiple sources and
multiple sinks).  In \cite{Yeung02first}, first steps were made to
characterize transmission throughput for the general case via
\emph{entropy functions} (polymatroids whose ground set is a set of
random variables, and whose rank function is Shannon entropy). Inner
and outer bounds on throughput were obtained in this way. Using the
same idea, an exact characterization of the set of all achievable
throughputs was later obtained~\cite{Yan2009The-Capacity}. Analogous
bounds for networks where intermediate nodes are restricted to use
only linear codes were obtained in~\cite{Chan07capacity} via
\emph{representable entropy functions}.

Unfortunately, these entropy function based characterizations are
implicit in nature, since an explicit characterization of the set of
all entropy functions is still missing. Characterization of this set
is one of the major open problems in information theory. Similarly,
the set of all representable entropy functions has no explicit
characterization. Notably, this set is a subset of representable
polymatroids, whose characterization is one of the major open problems
in matroid theory.

%However, by realizing that all all entropy functions are polymatroids, one can derive an explicit outer bound for the set of all achievable throughputs. The resulting bound is called \emph{Linear Programming (LP) bound}.  
%While existing work already pointed out that bounds and even the actual  set of achievable throughputs can be obtained from the set of entropy functions,

This lack of explicit, computable results could prompt one to question
this approach based on entropy functions. Although it leads to
attractive implicit characterizations, perhaps the difficulties that
arise are somehow an artefact of the approach. One could therefore be
tempted to seek simpler characterizations of transmission throughput
that avoid the need to precisely know the set of entropy functions.
For instance, \cite{Ahlswede.Cai.ea00network} demonstrated that a much
simpler characterization is possible in the single source scenario
where the max-flow min-cut bound is tight.  Unfortunately, a recent
paper~\cite{Chan2008Dualities} disproved the existence of any simpler
characterization for the general case. Using a specially contructed
network, it was proved that if one can determine the set of all
achievable throughputs in the special network, then one can also
determine the set of all entropy functions (and vice-versa). Hence,
determining achievable throughput for network coding is in general no
simpler than determining the set of entropy functions. A similar
duality was obtained in the same paper between the set of throughputs
achieved by linear codes and the set of representable entropy
functions.

These
results~\cite{Yeung02first,Yan2009The-Capacity,Chan07capacity,Chan2008Dualities}
indicate a very close tie between characterization of (representable)
entropy functions and throughput achievable with (linear) network
codes.  Characterization of entropy functions is equivalent to finding
all linear information inequalities that hold regardless of the
underlying joint distribution~\cite{Yeung02first}.  It is a well known
result, extending back to Shannon~\cite{Shannon48mathematical} that entropy and mutual
information are both nonnegative, correpsonding exactly to the
polymatroid axioms. No further information inequalities were found for
fifty years, until~\cite{Zhang.Yeung98characterization} reported the
first ``non-Shannon'' information inequality. The significance of that
result lay not only in the inequality itself, but also in its
construction. This particular approach for construction has been the
main ingredient in every non-Shannon inequality that has been
subsequently discovered. Using this appraoch, new inequalities can be
found mechanically~\cite{Dougherty.Freiling.ea06six} and there are in
fact infinitely many such independent inequalities even when there are
only four random variables involved~\cite{Matus07infinitely}. Despite
this progress, a complete characterization is still missing, and we
still only have one basic approach for finding new inequalities.
 
This situation does not improve for representable entropy
functions. In addition to the polymatroid inequalities, it is well
known that representable entropy functions satisfy Ingleton's
inequalities~\cite{ingleton71}.  Specifically, let
$\vecspace_1,\dots,\vecspace_4$ be vector subspaces. Then
\begin{multline}
  0 \le \dim \myspan{ \vecspace_{1},\vecspace_{2} } + \dim
  \myspan{\vecspace_{1},\vecspace_{3}}
  +\dim\myspan{\vecspace_{1},\vecspace_{4}} 
  + \dim\myspan{\vecspace_{2},\vecspace_{3}}
  +\dim\myspan{\vecspace_{2},\vecspace_{4}} \\ -\dim\myspan{\vecspace_{1}} 
  -\dim\myspan{\vecspace_{2}}
  -\dim\myspan{\vecspace_{3},\vecspace_{4}}
  -\dim\myspan{\vecspace_{1},\vecspace_{2},\vecspace_{3}}
  -\dim\myspan{\vecspace_{1},\vecspace_{2},\vecspace_{4}}
   \label{eqn:vecIngletonineq}
\end{multline}
where $\myspan{ \vecspace_{i},\vecspace_{j} }$ is the minimal vector
subspace containing $\vecspace_{i} \cup \vecspace_{j}$, and similar
for $\myspan{ \vecspace_{i},\vecspace_{j}, \vecspace_{k} }$.
%The inequalities \eqref{eqn:matroidIngletonineq} and
%\eqref{eqn:vecIngletonineq} are equivalent, via, $\dim \myspan{
%  \vecspace_{i}} = r(X_{i})$, $\dim \myspan{
%  \vecspace_{i},\vecspace_{j} } = r(X_{i},X_{j})$ and $\dim \myspan{
%  \vecspace_{i},\vecspace_{j},\vecspace_{k} } = r(X_{i},X_{j},X_k)$.
It has been an open problem since 1971 as to whether these
inequalities are also sufficient as well as being necessary conditions
for representability.

Very recently, several new inequalities for representable polymatroids
were reported at the 2009 Workshop on Applications of Matroid Theory
and Combinatorial Optimization to Information and Coding
Theory~\cite{Dougherty2009Non-Shannon}\footnote{We became aware of this independent work
  during the preparation of early drafts of this manuscript.}. These
inequalities were found by adapting the approach
in~\cite{Zhang.Yeung98characterization,Dougherty.Freiling.ea06six}. It
was verified numerically that the newly obtained inequalities (which
we shall refer to as \emph{DFZ inequalities}) completely characterize
representable entropy functions involving five variables (the Ingleton
inequalities are already known to be sufficient for four variables).
It is not known if these inequalities remain sufficient for more than
five variables.
 
The objective of this paper is to understand properties of
representable entropy functions (and more generally, representable
polymatroids).  Our main contribution is a proof for the insufficiency
of the Ingleton and DFZ inequalities for charcterization of
representable polymatroids.

Whereas~\cite{Dougherty2009Non-Shannon} constructively proves the insufficiency of the
Ingleton inequalities following the Zhang-Yeung
approach~\cite{Zhang.Yeung98characterization}, our approach is totally
different. We construct a polymatroid which satisfies every Ingleton
and DFZ inequality, but which is not contained within the minimal
closed and convex cone containing all representable polymatroids. This
directly establishes the existence of further, yet-to-be-discovered
linear inequalities for representable polymatroids.
 
The organization of the paper is as follows. In Section
\ref{sec:backbg}, we will introduce the required technical framework
for the problem. Section \ref{sec:newpoly} introduces a new method for
constructing a polymatroid by adhering together any two Ingleton
polymatroids. We will prove that the resulting polymatroids satisfy
the Ingleton inequalities, and that this construction also preserves
representability. In Section \ref{sec:constructed}, we will construct
an Ingleton polymatroid by adhering two \emph{representable}
polymatroids together.  This constructed polymatroid will be proved in
Section \ref{sec:main} to lie outside the closed and convex cone
containing all representable polymatroids. This establishes the
insufficiency of Ingleton's inequalities. Finally, in Section
\ref{sec:insufffive}, we prove that this constructed polymatroid also
satisfies the DFZ inequalities (for five variables), demonstrating the
insufficiency of the DFZ inequalities.

The following notational conventions will be used.  Set union will be
denoted by concatenation; Singletons and sets with one element are not
distinguished; Given $\X=\{X_1,X_2,\dots,X_n\}$ and any subset
$\alpha$ of the finite index set\footnote{If $n$ is understood, the
  subscript may be dropped for simplicity.} $\N_{n}=\{1,2,\dots,n\}$,
the subscript $X_\alpha$ will mean the set $\{X_i, i\in\alpha\}$. For
$\alpha,\beta\subseteq\N_{n}$, $X_{\alpha\beta} = X_\alpha X_\beta =
X_\alpha\cup X_\beta$ all refer to the same set.  Similarly, for any
$\A,\B \subseteq \X$, $\A \cup \B$ and $\A\B$ are the same set.  $\la
S \ra$ will denote the minimal vector space spanned by $S$.  We will
use $\con(\set{S})$ to denote the minimal convex cone containing the
set $\set{S}$ and $\barcon(\set{S})$ to denote the closure of
$\con(\set{S})$. Finally, $\reals$, $\integers$ and $\field_q$ are the reals,
integers and a finite field on $q$ elements.

%%%%%%%%%%%%%%%%%%%%%%%%%%%%%%%%%%%%%%%%%%%%
\section{Background}\label{sec:backbg}

A \emph{polymatroid} over the \emph{ground set}
$\X=\{X_1,X_2,\dots,X_n\}$ is a tuple $(\X, \bfh)$ where the
\emph{rank function} $\bfh:2^{\X}\mapsto\reals_+$ satisfies the
following axioms for all $\A,\B \subseteq \X$,
\begin{align}
  \bfh(\emptyset)  &= 0 \tag{R1} \label{eq:R1} \\
  \A\subseteq\B  &\implies \bfh(\A) \leq \bfh(\B)
  \tag{R2} \label{eq:R2}  \\
  \bfh(\A\cup\B) + \bfh(A\cap\B) &\leq \bfh(A) + \bfh(\B) \tag{R3}. \label{eq:R3} 
\end{align}
A polymatroid $(\X, \bfh)$ is called a \emph{matroid}~\cite{oxley92},
if it further satisfies the cardinality bound, $\bfh(\A)\leq|\A|$, and
the integrality constraint $\bfh(\A)\in\integers$, for all
$\A,\B\subseteq\X$.

For any $\A,\B, \C \subseteq\X$, define the generalized
information expressions as follows:
\begin{align}
  H\left(\A \mid \C \right) & \triangleq \bfh(\A\C)-\bfh(\C) \label{eq:conditionalentropy} \\
  I\left(\A;\B \mid \C\right) & \triangleq \bfh(\A \C)+\bfh(\B
  \C)
  -\bfh(\C)-\bfh(\A\B\C).\label{eq:conditionalmutualinformation}
\end{align}
when $\C=\emptyset$ we write $H(\A)=\bfh(\A)$ and
$I(\A;\B)=H(\A)-H(\A\mid\B)$ (consistent with the above definitions).
It is straightforward to prove that $(\X, \bfh)$ is a polymatroid if
and only if \eqref{eq:R1} holds and both \eqref{eq:conditionalentropy}
and \eqref{eq:conditionalmutualinformation} are nonnegative for all
choices of $\A,\B$ and $\set{C}$.

Polymatroids arise in many different contexts.  For example, let $\X =
\{ \seq X1n \}$ be a set of random variables. This naturally induces a
polymatroid $(\X, \bfh)$ such that $\bfh( \A )$ is the Shannon entropy
$H(\A)$ of the subset of random variables in $\A$.  In this case,
\eqref{eq:conditionalentropy} and
\eqref{eq:conditionalmutualinformation} are merely the usual
definitions for conditional entropy and mutual information, and $(\X,
\bfh)$ is a polymatroid due to the nonnegativity of (conditional)
entropies and mutual information. We emphasise however that the definitions
\eqref{eq:conditionalentropy} and
\eqref{eq:conditionalmutualinformation} are made for arbitrary
polymatroids (where the rank function may not induced by random
variables).

%We have already mentioned
%that~\eqref{eq:R1}-\eqref{eq:R3} are equivalent to the Shannon
%inequalities~\cite{Yeung02first}.  These notations
%reveal~\eqref{eq:R2} and \eqref{eq:R3} in a form more familiar to
%information theorists, namely
%\begin{align}\label{eq:shannonineq}
%H(X_{\alpha}|X_{\beta})\ge0 \text{ and } I(X_{\alpha};X_{\beta} |
%X_{\gamma}) \ge 0. 
%\end{align}

Polymatroids can also be induced by vector subspaces. 
Let  $\X  = ( \seq X1n)$ be a set  of  subspaces 
of a vector space $\mathbb W$ over a finite field $\field_{q}$. 
Define $\bfh(\A)$ as the dimension of the minimal vector subspace 
containing all the subspaces in $\A$,
\begin{align}\label{eq:vspacepoly}
\bfh(\A) \defined \dim \langle \A \rangle 
\end{align}
Then $(\X, \bfh)$ is also a polymatroid.  These subspace induced
polymatroids (called representable polymatroids) are of the main
objects of interest in this paper.

According to definitions \eqref{eq:conditionalentropy} and
\eqref{eq:conditionalmutualinformation} above, when $\bfh$ is defined
as in \eqref{eq:vspacepoly}, we can use $H(\A)$ to denote $\dim
\langle \A \rangle $ for any set of vector subspaces
$\A$. Furthermore, the following lemma may be easily verified.
\begin{lemma}
Let  $\X  = ( \seq X1n)$ be a set  of vector subspaces and 
 $\A,\B \subseteq \X$. Then 
\begin{align}
H(\A| \B) & = \dim \langle \A,\B \rangle - \dim \langle \B \rangle \\
I(\A ; \B) & = \dim  \left( \langle \A \rangle \cap  \langle \B \rangle  \right)
\end{align}
Furthermore, if $V = \langle \A \rangle \cap  \langle \B \rangle$, then 
$I(\A ; \B | V) = 0$.
\end{lemma}
We shall classify polymatroids as follows.
\begin{definitn}[Classification]\label{def:classification}
A polymatroid  $(\X,\bfh)$, and associated rank function $\bfh$, is called 
\begin{itemize}
\item \emph{$q$-representable} if there exists vector subspaces $\{
  \seq V1n \}$ over $\field_{q}$ such that for all
  $\alpha\subseteq\N_n$, $H(X_\alpha) = \dim \la V_\alpha \ra$ as
  defined in \eqref{eq:vspacepoly}.

\item \emph{representable} if is $q$-representable for some   $q$

\item \emph{even representable} if is $2^{m}$-representable for some
  positive integer $m$

\item \emph{odd representable} if is $p^{m}$-representable for some
  odd prime $p$ and a positive integer $m$

\item \emph{Ingletonian} if it satisfies the Ingleton inequality
  $J_{\bfh}(\setalpha_{1},\setalpha_{2},\setalpha_{3},\setalpha_{4})
  \ge 0 $ for all subsets $\setalpha_{1},\setalpha_{2},\setalpha_{3},
  \setalpha_{4}\subseteq\X$ where
   \begin{multline}\label{eqn:matroidIngletonineq}
     J_{\bfh}\left(\A_1,\A_2,\A_3,\A_4\right) \defined\ \bfh(\A_{1}
     \A_{2})+\bfh(\A_{1} \A_{3})+\bfh(\A_{1}\A_{4}) +\bfh(\A_{2}
     \A_{3}) +\bfh(\A_{2} \A_{4})\\ -\bfh(\A_{1}) -\bfh(\A_{2})
     -\bfh(\A_{3} \A_{4}) -\bfh(\A_{1} \A_{2} \A_{3}) -\bfh(\A_{1}
     \A_{2} \A_{4}).
  \end{multline}
  
%\item[(P)] \emph{Polymatroidal} 
% if  \eqref{eq:conditionalentropy} and
% \eqref{eq:conditionalmutualinformation} are nonnegative for all
% $\alpha,\beta,\gamma\subseteq \N$. 
 
\end{itemize}
\end{definitn}

According to \cite{ingleton71}, if a polymatroid $(\X,\bfh)$ is
representable, then it is also Ingletonian. A natural question then
arises -- \emph{Are the Ingleton inequalities sufficient to
  characterize representable polymatroids?} In this paper, we will
show that the answer to this question is negative.
 
Following the framework for information inequalities presented
in~\cite{Yeung97framework}, it is useful to treat a rank function as a
vector or point in a $2^{|\X|}$-dimensional real Euclidean space whose
coordinates are indexed by the power set of $\X$.  Thus a point
$\bfh\in\Real^{2^{|\X|}}$ is specified by its coordinates as $\bfh =
(\bfh(\A) :\A \subseteq\X)$.  While a polymatroid is defined by a
ground set and a rank function, the ground set is actually implicitly
defined by the rank function. Hence, strictly speaking, a polymatroid
$(\X,\bfh)$ is no more than a rank function that satisfies the
polymatroidal axioms. In other words, a polymatroid is merely a point
in an Euclidean space and characterization of representable
polymatroids is equivalent to characterizing those points induced by
representable polymatroids.

Regarding polymatroids as points in $\Real^{2^{|\X|}}$ permits us to
define metrics and limits on the set of polymatroids. Let
$\Upsilon_{q}[\X]$ be the set of all $q$-representable rank functions
and $\Upsilon[\X]= \bigcup_{q} \Upsilon_{q}[\X])$ be the set of all
representable rank functions.
\begin{definitn}
  A polymatroid $(\X, \bfh)$ (and the corresponding rank function
  $\bfh$) is called \emph{almost representable} if there exists a
  sequence of representable rank functions
  $\{\bfg_{i}\}_{i=1}^{\infty}$ and a sequence of positive numbers
  $c_{i}$ such that $\bfh = \lim_{i\to\infty} c_{i} \bfg_{i}$.  On the
  other hand, $\bfh$ is called \emph{cc-representable}\footnote{
    ``cc'' is a mnemonic for ``Closed and Convex cone''.}  if $\bfh
  \in \barcon(\Upsilon[\X])$.
\end{definitn}

A linear inequality involving polymatroids is merely a linear
inequality over $\Real^{2^{|\X|}}$.  We are interested to determine
necessary conditions (linear inequalities in particular) on the rank
function $\bfh$ under which it is representable. In this paper,
complete characterization of representable rank functions means an
explicit determination of $\barcon(\Upsilon[\X])$ or the set of linear
inequalities satisfied by points in $\barcon(\Upsilon[\X])$. The
following proposition may be directly verified.
\begin{prop}
  A linear inequality $\sum_{i} c_{i} H(\A) \ge 0$ holds for all
  representable polymatroids (i.e. is a valid subspace rank
  inequality) if and only if $\sum_{i} c_{i} \bfh(\A) \ge 0$ for all
  $\bfh \in \barcon(\Upsilon [\X])$.
\end{prop}

%%%%%%%%%%%%%%%%%%%%%%%%%%%%%%%%%%
\def\define{\triangleq}
\def\p{\prime}
\def\Y{{\mathcal Y}}
\def\Ye{{\mathcal Y}^{\epsilon}}
\def\Z{{\mathcal Z}}
\def\da{{\times_{d}\:}}

%%%%%%%%%%%%%%%%%%%%%%%%%%%%%%%%%%
\section{Creating new polymatroids}\label{sec:newpoly}
We now propose an approach to perturb an Ingletonian polymatroid in a
way that preserves the Ingletonian property. This is achieved by
Theorem~\ref{thm:extension}. We shall subsequently show in
Theorems~\ref{thm:limitperturbone} and~\ref{thm:perturbrepresentable}
that this perturbation also preserves (almost) representability.  In
Section \ref{sec:main}, we will use this approach to perturb a member
of $\barcon(\Upsilon[\X])$, taking it outside of
$\barcon(\Upsilon[\X])$. This perturbed polymatroid will be used to
show the existence of new subspace rank inequalities for representable
polymatroids.

\begin{thm}[$\epsilon$-pertubation]\label{thm:extension}
  Let $(\Y, \bfh)$ be an Ingletonian polymatroid. Let $0 \le \epsilon
  \le \bfh(\Y)$ and define for all $\A \subseteq \Y$
  \begin{align}\label{eqn:extension}
    \bfg(\A ) &\define \min ( \bfh(\A ) ,  \bfh(\Y) - \epsilon ).
  \end{align}
  Then $(\Y, \bfg)$ is also an Inlgetonian polymatroid.

\end{thm}
\begin{proof}
We need to prove that  
\begin{align}\label{eqn:ineq}
J_{\bfg}\left(\V_1, \V_2,\V_3,\V_4\right)  \ge 0
\end{align}
for all $\seq \V14\subseteq\Y$. To simplify notation, define
\begin{align*}
  J^{+}_{\bfg}\left(\V_1,\V_2,\V_3,\V_4\right) &\defined \bfg(\V_1\V_2)
  + \bfg (\V_1\V_3) +\bfg (\V_1\V_4) + \bfg (\V_2\V_3) + \bfg
  (\V_2\V_4)\\
  J^{-}_{\bfg} \left(\V_1,\V_2,\V_3,\V_4\right) &\defined \bfg
  (\V_1)+\bfg (\V_2) + \bfg (\V_3\V_4) + \bfg (\V_1\V_2\V_3) + \bfg
  (\V_1\V_2\V_4).
\end{align*}
With these definitions, the Ingleton inequality \eqref{eqn:ineq} is
written
\begin{align}\label{eq:star}
J^{+}_{\bfg}\left(\V_1,\V_2,\V_3,\V_4\right) \ge J^{-}_{\bfg}\left(\V_1,\V_2,\V_3,\V_4\right).
\end{align}
It is straightforward to prove that $(\Y, \bfg)$ is a polymatroid. We
must additionally show that it is Ingletonian. Let
 \[
 U = \{ (1,2), (1,3), (2,3), (1,4), (2,4) \}.
 \] 
 be the collection of $\alpha \subseteq \N_{4}$ such that the summand
 $\bfg(\V_{i},i\in\alpha)$ appears in
 $J^{+}_{\bfg}\left(\V_1,\V_2,\V_3,\V_4\right) $.  Let $Q \define \{
 \alpha \subseteq \N_{4}: \bfh(\Y) - \epsilon \le \bfh(\V_i,
 i\in{\alpha}) \}$. Thus $Q$ identifies summands in $J^+$ and $J^-$
 for which the $\epsilon$-perturbation in~\eqref{eqn:extension} bites.
Note that if $\alpha \in Q$, then $\bfg(\V_i,
 i\in\alpha) \ge \bfg(\V_i, i\in{\beta})$ for all $\beta \subseteq
 \N_{4}$. 

 We will now proceed on a case-by-case basis, proving that
 \eqref{eqn:ineq} holds in the following distinct and exhaustive
 cases.

\noindent
\textbf{Case 1}: $Q \cap U=\emptyset$.

Inequality \eqref{eqn:ineq} clearly holds and follows from the
fact that $(\Y, \bfh)$ is Ingletonian, $J^{+}_{\bfg} = J^{+}_{\bfh}$
and $J^{-}_{\bfg} \le J^{-}_{\bfh}$.

\noindent
\textbf{Case 2}: $Q \cap U =\{ (1,2) \}$. 

In this case, $\bfh(\V_1\V_2) \ge \bfh(\Y)-\epsilon$. By monotonicity
of polymatroids~\eqref{eq:R2}, $\bfh(\V_1\V_2\V_{3}) \ge
\bfh(\Y)-\epsilon$.  The left hand side of \eqref{eq:star} thus
becomes
\[
J^{+}_{\bfg}\left(\V_1,\V_2,\V_3,\V_4\right) = J^{+}_{\bfh}\left(\V_1,\V_2,\V_3,\V_4\right)   -  \bfh(\V_1\V_2) + \bfh(\Y) - \epsilon.
\]
Similarly, its right hand side can be shown to be bounded above by    
\[
J^{-}_{\bfg}\left(\V_1,\V_2,\V_3,\V_4\right) = J^{-}_{\bfh}\left(\V_1,\V_2,\V_3,\V_4\right)  -  \bfh(\V_1\V_2\V_{3}) + \bfh(\Y) - \epsilon.
\]
Thus \eqref{eqn:ineq} holds, since $(\Y,\bfh)$ is Ingletonian and
$\bfh(\V_1\V_2) \le \bfh(\V_1\V_2\V_3)$. A similar approach may be
used when $Q \cap U$ is any one of $(1,3) $, $(1,4)$, $(2,3)$ or $(2,4) $.

\noindent
\textbf{Case 3}: $Q \cap U = \{ (1,3) , (2,3) \}$. 

As $\bfg$ is a polymatroid,
$\bfg(\V_1\V_2)+\bfg(\V_1\V_4)+\bfg(\V_2\V_4) \ge \bfg(\V_{1}) +
\bfg(\V_2) + \bfg(\V_1\V_2\V_4)$ and hence \eqref{eqn:ineq} holds as
$\bfg(\V_1\V_3)+\bfg(\V_2\V_3) = 2\bfg(\Y) \ge
\bfg(\V_3\V_{4})+\bfg(\V_1\V_2\V_3)$.  Similarly, \eqref{eqn:ineq}
holds if $Q \cap U = \{ (1,4) , (2,4) \}$.

%%%%
\noindent
\textbf{Case 4}: $Q \cap U = \{(1,2) , (1,3) \} $.

Again, $\bfg(\V_2\V_3)+\bfg(\V_1\V_4)+\bfg(\V_2\V_4) \ge \bfg(\V_1) +
\bfg(\V_2) + \bfg(\V_3\V_4)$ and consequently \eqref{eqn:ineq} holds
because $\bfg$ is a polymatroid.
Using the same argument, \eqref{eqn:ineq} also holds when $Q \cap U$
is either $\{(1,2) , (1,4) \} $, $\{(1,2) , (2,3) \} $ or $\{(1,2) ,
(2,4) \}$.

\noindent
\textbf{Case 5}: $Q \cap U = \{(1,3) , (2,4) \} $. 

Now $\bfg$ is a polymatroid and hence
$\bfg(\V_1\V_2)+\bfg(\V_2\V_3)+\bfg(\V_1\V_4) \ge  \bfg(\V_1) + \bfg(\V_2) + \bfg(\V_1\V_2\V_3)$.
Consequently, \eqref{eqn:ineq}  holds as in the previous case. 
Similarly, \eqref{eqn:ineq}  holds when $Q \cap U = \{(1,4) , (2,3) \} $.

\noindent
\textbf{Case 6}: $Q \cap U = \{(1,3) , (1,4) \} $.

In this case, \eqref{eqn:ineq} holds because 
$\bfg(\V_1\V_2)+\bfg(\V_2\V_3)+\bfg(\V_2\V_4) \ge  \bfg(\V_1) + \bfg(\V_2) + \bfg(\V_2\V_3\V_4)$.

\noindent
\textbf{Case 7}: $ | Q\cap U | = 3$ and $Q\cap U = \{ (1,2), (2,3), (2,4)  \}$. 

Inequality \eqref{eqn:ineq} follows from
\begin{align*}
\bfg(\V_1\V_3)+\bfg(\V_1\V_4) &\ge  \bfg(\V_1) + \bfg(\V_1\V_3\V_4) \\
&\ge   \bfg(\V_1) + \bfg(\V_3\V_4).
\end{align*}
A similar approach can be used for other cases when $| Q\cap U |
\ge 3$.
\end{proof}

In Theorem \ref{thm:extension}, we proved that the
$\epsilon$-perturbation of an Ingletonian polymatroid is also
Ingletonian.  Theorem~\ref{thm:limitperturbone} shows that
$\epsilon$-perturbation also preserves representability.
\begin{thm}\label{thm:limitperturbone}
  Suppose $(\X,\bfh)$ is representable and let $(\X,\bfg_{\epsilon})$ be
  its $\epsilon$-perturbation \eqref{eqn:extension}. 
%In other words, 
%\begin{align}
%\bfg_{\epsilon}(\A ) &\define \min ( \bfh(\A ) ,  \bfh(\Y) - \epsilon )
%\end{align}
  Then $(\X,\bfg_{\epsilon} )$ is almost representable for any $0 \le
  \epsilon \le \bfh(\X)$. It is also representable if 
  $\epsilon\in\integers$.
\end{thm}
Before we prove Theorem \ref{thm:limitperturbone}, we require some
basic results regarding vector subspaces.  Let $\bb{A}$ be a vector
subspace of $\bb{W}$. Define $\bb{A}^{*}$ as a subspace of $\bb{W}$
such that
\begin{enumerate}
\item
$\la\bb{A},\bb{A}^{*}\ra=\bb{W}$, and hence $H(A,A^{*}) = H(W)$
\item 
$\bb{A} \cap \bb{A}^{*} = \{{\bf 0}\}$. 
\end{enumerate} 
Clearly, any vector $\bf u\in\bb{W}$ can be written uniquely as ${\bf
  u}={\bf u}_1+{\bf u}_2$ where ${\bf u}_{1}\in \bb{A}^{*}$ and $ {\bf
  u}_{2}\in \bb{A}$.  We will refer to ${\bf u}_{1}\defined T_\bb{A}
({\bf u})$ as the projection of $\bf u$ away from $\bb{A}$.
While $T_\bb{A} ({\bf u})$ depends on the choice of $\bb{A}^{*}$, it
can be directly verified that the following lemma holds for all
legitimate choices of $\bb{A}^{*}$.
\begin{lemma} \label{lemma:property}
Let $\bb{B}$ and $\bb{C}$ be subspaces of $\bb{W}$. Then for any subspace $A$ of $W$, 
\[
 T_{\bb{A}}(\bb{B}) \defined \{  T_{\bb{A}}({\bf u}) : \: {\bf u} \in \bb{B}  \}
 \]
  is a subspace of $\bb{W}$. Furthermore,
\begin{align}
H\left(T_{\bb{A}}(\bb{B}) \right) &= H\left(B\mid A\right),\\
H\left(T_{A}( \la B_{j}, j\in\beta \ra) \right)  &= H\left(T_{A}( B_{j}) , j\in\beta  \right).
\end{align}
Consequently, 
\begin{enumerate}
\item $ H\left(\bb{B}_{j} , j\in\beta \right) \ge  H\left(T_{\bb{A}}(\bb{B}_{j}) , j\in\beta\right)  \ge H\left(\bb{B}_{j}, j\in\beta\right) -  H\left(\bb{A}\right)$ 
\item 
If  $H\left(B\mid C\right) = 0$ (i.e. $\bb{B} \subseteq \bb{C}$), then $H\left(T_{\bb{A}}(\bb{B}) \mid T_{\bb{A}}(\bb{C})\right) = 0$. 
More generally, if $H\left(B \mid C_{i}, i\in\alpha\right) = 0$, then $H\left(T_{\bb{A}}(B) \mid T_{\bb{A}}(C_{i}), i\in\alpha\right)=0$.  

\item If $\bb{B} \cap \bb{A} = \{ \bf 0 \}$ (i.e. $I\left(A;B\right)=0$), then $H\left(\bb{B}\right) = H\left(T_{\bb{A}}(\bb{B})\right)$.

\end{enumerate}  
\end{lemma}
The projection operator $T_{A}( \cdot )$ has a natural
interpretation. Specifically, let $\{\seq X1n, A \}$ be a collection
of subspaces in $W$, which induces a representable polymatroid in the
usual way. % via
% \begin{align*}
% H\left(X_i, i\in\alpha\right) & \defined \dim \la X_i, i\in\alpha \ra \\
% H\left(A,X_i, i\in\alpha\right) & \defined \dim \la A, X_i, i\in\alpha \ra. 
% \end{align*}
By Lemma \ref{lemma:property}, there exists subspaces $\{\seq B1n\}$ of $W$ such that $B_{i} \defined T_{A}(X_{i})$ and 
\begin{align}
H\left(B_{i}, i\in\alpha\right) = H\left(X_{i},i\in\alpha \mid A\right).
\end{align} 
In other words, using the projection operator $T_{A}( \cdot )$, one
can transform any set of subspaces $\{\seq X1n \}$ into another
set $\{\seq B1n \}$ such that $H\left(B_{i}, i\in\alpha\right) =
H(X_{i},i\in\alpha | A)$.
\begin{proof}[Proof of Theorem \ref{thm:limitperturbone}]
  Suppose $(\X,\bfh)$ is representable. If $\bfh(\X) = 0$, then the
  theorem is trivial. Now suppose $\bfh(\X) > 0 $ and hence $\bfh(\X)
  \ge 1$ (since it is the dimension of a space).  We  begin by
  proving that $(\X,\bfg_{\epsilon})$ is representable (and hence
  almost representable) when $\epsilon=1$. The argument will
  subsequently be extended to cover other values of $\epsilon$.

  First, as $(\X,\bfh)$ is representable, there exists subspaces
  $V_{i}$, $i\in\N$ such that
  \begin{align}
    \bfh(X_{i},i\in\alpha) = H(V_{i},i\in\alpha) \defined 
    \dim \la V_{i},i\in\alpha \ra.
  \end{align}
  Assume without loss of generality that these subspaces are over an
  underlying field $\field_{q}$. For any positive integer $m$,
  $\field_{q}$ can be regarded as a subfield of $\field_{q^{m}}$. In
  fact, for each $i$, let $V_{i}^{(m)}$  be the subspaces over $\field_{q^{m}}$ 
  spanned by $V_{i}$. Then 
    \begin{align}
    \dim \la V_{i}^{(m)}, i\in\alpha \ra = \dim \la V_{i},  i\in\alpha \ra  
  \end{align}
  or equivalently, $H( V_{i}^{(m)}, i\in\alpha) = H(V_{i},  i\in\alpha) $.

  Let $C= H(V_{\N})$ and $Q \defined \{ \alpha \subseteq \N : \:
  H(V_{\alpha}) < C \}$. Then for each $\alpha \in Q$,
  \[
  H(V_{\alpha}^{(m)}) = H(V_{\alpha}) \le C -1.
  \]
  The volume (i.e. cardinality) of $ \la V_{i}^{(m)}, i\in\alpha \ra$
  is at most $(q^{m})^{C-1}$ while the volume of $ \la V_{i}^{(m)}, i
  \in\N\ra$ is $(q^{m})^{C}$. Hence, for sufficiently large $m$,
  $\bigcup_{\alpha\in Q} V_{i}^{(m)} $ is a proper subset of $ \la
  V_{i}^{(m)}, i \in\N\ra$.

  Let ${\bf u} \in \la V_{i}^{(m)}, i \in\N\ra$ but not in
  $\bigcup_{\alpha\in Q} V_{i}^{(m)} $. Let $A=\la{\bf u}\ra$ and define
  \begin{align}
    B_{i} \defined T_{A}(V_{i}^{(m)}).
  \end{align}
  By Lemma \ref{lemma:property}, it is straightforward to prove that
  $\bfg_{\epsilon}(\alpha) = H(B_{i},i\in\alpha)$. Hence,
  $\bfg_{\epsilon}$ is representable.  Repeating the same argument
  multiple times, we can also prove that the theorem also holds when
  $\epsilon$ is a positive integer.
%In the extreme, the theorem holds when $\epsilon = h(\X)$ and thus

  Now, suppose $\epsilon=k/\ell$ is rational.  For any representable
  $(\X,\bfh)$, it is easy to find another representable $(\X,\bff)$
  such that $\bff = \ell\bfh $. Consequently
\begin{align}
\bfg_{\epsilon}  (\A) &= \min ( \bfh (\A),  \bfh (\X)-k/\ell) \\
&= \frac{1}{\ell} \min ( \ell\bfh (\A),  \ell\bfh (\X)-k) \\
&= \frac{1}{\ell} \min (  \bff (\A),  \bff (\X)-k).
\end{align}
is almost representable.  

Finally, the remaining case when $\epsilon$ is irrational can be
proved by a continuity argument. Specifically, let $\nu_{j}$ be a
sequence of rational numbers converging to $\epsilon$, then it is easy
to prove that $\lim_{j\to\infty} \bfg_{\nu_{j}} =
\bfg_{\epsilon}$. Hence, $\bfg_{\epsilon}$ is almost representable.
\end{proof}

The final result of this Section,
Theorem~\ref{thm:perturbrepresentable} is a direct consequence of the
following proposition.
\begin{prop}\label{prop:perturbrepresentable}
  Let $\{(\X, \bfh_{i})\}_{i=1}^{\infty}$ be a sequence of
  representable polymatroids such that $\lim_{i\to\infty} c_{i}
  \bfh_{i} = \bfh_{0}$ for some positive sequence of numbers
  $\{c_{i}\}_{i=1}^{\infty}$.  Suppose $0 \le \epsilon \le
  \bfh_{0}(\X)$.  For each $i$, define
\begin{align}
\bfg_{i} (\A) &\defined \min (c_{i} \bfh_{i}(\A), c_{i}\bfh_{i}(\X)-\epsilon) \\
\bfg_{0}  (\A) &\defined \min ( \bfh_{0}(\A),  \bfh_{0}(\X)-\epsilon). 
\end{align}
Then $\lim_{i\to\infty} \bfg_{i} = \bfg_{0}$ and $\bfg_{0}$ is
almost representable.
\end{prop}
% Proof
\begin{proof}
  If $\epsilon = \bfh_{0}(\X)$, then the proposition is obvious. Now,
  suppose $\epsilon < \bfh_{0}(\X)$.  By the continuity of
  $\min(a,b)$,
\[
\lim_{i\to\infty} \left( \min \left(c_{i}\bfh_{i}(\A), c_{i}\bfh_{i}(\X)-\epsilon\right)  \right) = \min \left(\lim_{i\to\infty} c_{i}\bfh_{i}(\A), \lim_{i\to\infty} c_{i}\bfh_{i}(\X)-\epsilon\right) 
\]
and hence $\lim_{i\to\infty} \bfg_{i} = \bfg_{0}$. 
%notice that 
%\begin{align}
%\lim_{i\to\infty}\bfg_{i} (\A) & = \lim_{i\to\infty} \left( \min (c_{i}\bfh_{i}(\A), c_{i}\bfh_{i}(\X)-\epsilon)  \right) \\
%& =   \min (\lim_{i\to\infty} c_{i}\bfh_{i}(\A), \lim_{i\to\infty} c_{i}\bfh_{i}(\X)-\epsilon)  \\
%& = \min (\bfh_{0}(\A), \bfh_{0}(\X)-\epsilon) \\
%& = \bfg_{0}(\A).
%\end{align}
 
On the other hand, since $\lim_{i\to\infty}c_{i} \bfh_{i} = \bfh_{0}$
and $0 \le \epsilon < \bfh_{0}(\X)$, we have for sufficiently large
$i$ that $0 \le \epsilon \le c_{i}\bfh_{i}(\X) $ or equivalently $0
\le \epsilon/c_{i} \le \bfh_{i}(\X) $.  As $\bfh_{i}$ is
representable, $\bfg_{i} / c_{i} $ and hence $\bfg_{i}$ are almost
representable by Theorem \ref{thm:limitperturbone}.  Consequently, its
limit $ \bfg_{0}$ is also almost representable.
\end{proof}
\begin{thm}\label{thm:perturbrepresentable} 
Suppose $(\X,\bfh)$ is almost representable. Let $(\X,\bfg)$ be its $\epsilon$-perturbed polymatroid as defined in \eqref{eqn:extension}. Then  $\bfg$ is almost representable.  
\end{thm}

\section{Main results} \label{sec:main} The main result of this paper
is the following theorem, which is a direct consequence of
Theorem~\ref{thm:main}, which we prove in this section, and
Theorem~\ref{thm:existence} in the following section (which
establishes the existence of certain matroids required for
Theorem~\ref{thm:main}).
\begin{thm}[Insufficiency of Ingleton's inequalities]\label{th:notccrep} 
  There exists an Ingletonian polymatroid that is not
  cc-representable.  Consequently, there are linear inequalities
  satisfied by all representable polymatroids but not implied by
  Ingleton's inequalities.
\end{thm}

So far, we have defined the concept of $\epsilon$-perturbation and
proved that it preserves both the Ingletonian property and
representability.  We will now use $\epsilon$-perturbation to
construct an Ingletonian polymatroid that is not cc-representable.
First, we will need to establish a key lemma concerning
\emph{connected matroids}~\cite[Chapter 4]{oxley92}.
\begin{definitn}[Connected Matroid]
  A matroid $(\X, \bfh)$ is \emph{connected} if for any pair of
  $X_{1}$ and $X_{2}$ in $\X$, there exists a circuit contains both
  $X_{1}$ and $X_{2}$.
\end{definitn} 
\begin{definitn}
  Let $M=(\X,\bfh)$ be a matroid. Define $\set{I}(M)$ as the set of
  equalities of the forms $H\left(\A\mid\B\right)=0$  or  
  $I\left(\A;\B\right) = 0$ satisfied by $M$.
\end{definitn}

%\begin{definitn}[Circuit graph]
%Consider a matroid $(\X, \bfh)$. It will induce an undirected \emph{circuit graph} such that its set of vertices is $\X$ and that two vertices $X_{1}$ and $X_{2}$ are  adjacent to each other if there exists a  circuit in $(\X, \bfh)$ that contains $X_{1}$ and $X_{2}$.
%%Let $X_{1},X_{2} \in \X$. We define a binary relation 
%%$X_{1} \sim X_{2}$ if there exists a  circuit in $(\X, \bfh)$ that contains $X_{1}$ and $X_{2}$. 
%%Furthermore, $X_{1} \sim_{c} X_{2}$ if there exists a sequence of 
%A matroid $(\X, \bfh)$ is called \emph{connected} if its induced circuit graph has only one connected component. 
%\end{definitn}

\begin{lemma}\label{lemma:circuit}
  If a matroid $M=(\X, \bfh)$ is connected and $(\X, \bfg)$ is any
  polymatroid that satisfies the set of $\set{I}(M)$ equalities
  induced by $M$, then $\bfg(X_{i})$ is constant for all $X_{i} \in
  \X$ and $\bfg = c \bfh$ for some $c\ge 0$.
\end{lemma}
\begin{proof}
  Suppose $X_{1}, X_{2}$ belong to a circuit in the matroid $M$. Then
  there exists a subset of variables $\A$ such that
  \begin{align}\label{eq:circuit}
    H(X_{1}| X_{2}, \A) = H(X_{2}| X_{1}, \A)  = I (X_{1} ; \A) =  I (X_{2} ; \A) =0
  \end{align}
  where \eqref{eq:circuit} is with respect to $\bfh$.  By assumption,
  $\bfg$ satisfies $\set{I}(M)$, which includes \eqref{eq:circuit}.
  It is easy to prove that if $(\X, \bfg)$ is a polymatroid also
  satisfying \eqref{eq:circuit}, then $\bfg(X_{1}) = \bfg(X_{2})$.  By
  the connectedness of $M$, $\bfg(X_{i})$ is constant for all $X_{i}
  \in \X$.

  Now, let $\B$ be any subset of $\X$. Since $(\X, \bfh)$ is a
  matroid, there exists $\A\subset\B$ such that $\bfh(\B|\A)=0$ and
  $\bfh(\A) = \sum_{X_{i} \in \A} \bfh(X_{i})$, and these identities
  belong to $\set{I}(M)$.  By assumption, $(\X, \bfg)$ also satisfies
  $\set{I}(M)$, and hence $\bfg(\B|\A)=0$ and $\bfg(\A) = \sum_{X_{i}
    \in \A} \bfg(X_{i})$. Thus $\bfh(\B) = | \A |\, \bfh(X_{i})$ and
  $\bfg(\B)= | \A |\, \bfg(X_{i})$.
\end{proof}

The main result of this paper hinges on the following theorem. It
describes an approach to adhere two representable matroids together in
such a way that the resulting polymatroid is not cc-representable. We
establish this theorem for connected matroids that are even (and odd)
representable but not almost odd (even) representable. The existence
of such \emph{strictly even} and \emph{strictly odd} matroids will
be established in Theorem \ref{thm:existence}. Together with Theorem
\ref{thm:main}, this provides the proof of Theorem~\ref{th:notccrep},
namely the insufficiency of the Ingleton inequalities .
\begin{thm}\label{thm:main}
  Let $(\X_{1}, \Phi_{1})$ (and $(\X_{2}, \Phi_{2})$) be an even (and
  odd) connected representable matroid that is not almost odd (almost
  even) representable.  Let $(\X,\Phi)$ be the direct sum of these two
  matroids, namely $\X = \X_{1} \cup \X_{2}$ and $\Phi(\A) =
  \Phi_{1}(\A \cap \X_{1}) + \Phi_{2}(\A \cap \X_{2})$ for all $\A
  \subseteq \X$.  Suppose $0 < \epsilon \le \min (\Phi_{1}(\X_{1} ),
  \Phi_{2}(\X_{2} )) $. Then $(\X, \Phi^{\epsilon})$ is not
  cc-representable.
\end{thm}
\begin{proof} 
  By definition of $\epsilon$-perturbation \eqref{eqn:extension}, it
  is easily verified that
  \begin{align}
    \Phi^{\epsilon}(\A) &= \Phi_{1}(\A), \quad \forall  \A \subseteq \X_{1} \\
    \Phi^{\epsilon}(\B) &= \Phi_{2}(\B), \quad \forall  \B \subseteq \X_{2} .
  \end{align}
  Hence, $\Phi^{\epsilon}$ satisfies all the equalities
  $\set{I}(M_{1})$ and $\set{I}(M_{2})$.

  Suppose to the contrary that
  $\Phi^{\epsilon}\in\barcon(\Upsilon[\X])$.  Then by definition there
  exists a sequence of points $\bfh_{i}\in\con(\Upsilon[\X])$ such
  that $\lim_{i\to\infty}\bfh_{i} = \Phi^{\epsilon}$.  As each
  $\bfh_{i}$ is a point in a $ 2^{|\X|}$-dimensional Euclidean space,
  Caratheodory's theorem allows each $\bfh_{i}$ to be written
  \begin{align}
    \bfh_{i} = \sum_{j=1}^{2^{|\X|}+1} c_{i,j}  \bff_{i,j} 
  \end{align}
  where $c_{i,j} \ge 0$ and $d_{i,j} \bff_{i,j}$ is representable for
  some $d_{i,j}>0$.
 
  Assume without loss of generality that $\bfh_{i}(\X)=\bff_{i,j}(\X)
  =\Phi^{\epsilon}(\X) $ for all $i,j$.  Then all $\bff_{i,j}$ are
  contained in the compact set $\{ \bff \in \Real^{2^{|\X|}} : 0 \le
  \bff(\A) \le \Phi^{\epsilon}(\X) \text{ for all } \A \subseteq \X
  \}$ and $\sum_{j=1}^{2^{|\X|}+1} c_{i,j} = 1$ (and hence $0 \le
  c_{i,j} \le 1$).  According to the Bolzano-Weierstrass theorem, any
  bounded sequence in a finite dimensional Euclidean space has a
  convergent subsequence. We may therefore assume without loss of
  generality the existence of the following limits for any
  $j=1,\ldots, 2^{|\X|}+1$
  \begin{align*}
    \lim_{i\to\infty} c_{i,j} &= c_{j} \\
    \lim_{i\to\infty} \bff_{i,j} &= \bff_{j}.
  \end{align*}
  Hence $\Phi^{\epsilon} = \sum_{j=1}^{2^{|\X|}+1} c_{j}\bff_{j}. $

  As each $\bff_{j}$ is the limit of a sequence of polymatroids
  $\{\bff_{i,j}\}_{i=1}^{\infty}$, $(\X , \bff_{j})$ is also a
  polymatroid for all $j$. Therefore, $(\X , \bff_{j})$ also satisfies
  equalities $\set{I}(M_{1})$ and $\set{I}(M_{2})$.  On the other
  hand, as $\epsilon > 0$, $ \Phi^{\epsilon}( \X_{1} ) +
  \Phi^{\epsilon}( \X_{2} ) > \Phi^{\epsilon}( \X) $, there is at
  least one $j$ such that
  \[
  \bff_{j}( \X_{1} ) + \bff_{j}( \X_{2} ) >  \bff_{j} ( \X ).
  \]
  Consequently, both $ \bff_{j}( \X_{1} )$ and $\bff_{j}( \X_{2} ) $
  are positive as $\bff_{j}$ is a polymatroid. Furthermore, since
  $(\X_{1}, \Phi_{1})$ and $(\X_{2}, \Phi_{2})$ are connected
  matroids, by Lemma \ref{lemma:circuit}, there exists positive
  constants $c$ and $c^{\p}$ such that
  \begin{align}
    \bff_{j}(\A) & = c \Phi_{1}(\A), \quad \forall  \A \subseteq \X_{1}, \label{eqn:matroida} \\
    \bff_{j}(\B) & = c^{\p} \Phi_{2}(\B), \quad \forall  \B \subseteq \X_{2}. \label{eqn:matroidb}
\end{align}
 
So far, we have proved that if $\Phi^{\epsilon} \in
\barcon(\Upsilon[\X])$, then there exists a sequence of polymatroids
$\{ (\X, \bff_{i,j} )\}_{i=1}^{\infty}$ such that (1) $d_{i,j}
\bff_{i,j} $ is representable for some $d_{i,j} > 0$, and (2) its
limit $\bff_{j}$ satisfies \eqref{eqn:matroida} and
\eqref{eqn:matroidb}.
%
%Let $\bfg_{i} = d_{i,j} \bff_{i,j}$ and $d_{i} = 1/ d_{i,j}$. 
%As $\bfg_{i}$ is representable, we can obtain a subsequence of which such that all the $\bfg_{i}$ in the subsequence are either all even or odd representable.
We may further assume without loss of generality that $d_{i,j}
\bff_{i,j} $ is either even representable or odd representable for all
$i$. Suppose first that all $d_{i,j} \bff_{i,j} $ are even
representable for all $i$.  The fact that $\lim_{i\to\infty}
\bff_{i,j}(\B) = \bff_{j}(\B) = c^{\p} \Phi_{2}(\B)$ thus implies that
$(\X, \Phi_{2})$ is almost even representable, contradicting the
hypothesis. Similarly, contradiction occurs if $d_{i,j} \bff_{i,j} $
are odd representable for all $i$. Contradiction occurs in both cases
and hence the theorem is proved.
 \end{proof}
%%%%%%%%%%%%%%%%%%%%%%%%%%%%%%%%%%%%%%
\section{Strictly Odd and Even Matroids}\label{sec:constructed}
\newcommand{\fn}[2]{{H_{\bff_{#1}} ( #2) }} In this section, we will
construct two representable matroids $(\X_{1}, \Phi_{1})$ and $(\X_{2}
, \Phi_{2})$ that satisfy the conditions given in Theorem
\ref{thm:main}. These matroids correspond to the first and second
networks in~\cite[Section II]{Dougherty.Freiling.ea05insufficiency}.

\def\u{{\bf u}}
 
Define the matroid $(\X_{1},\Phi_{1})$ with $\Phi_1(\cdot)=\dim\la\cdot\ra$
and ground set
\begin{align}\label{eq:firstd}
\X_{1} &\defined \{Y_{1},Y_{2},Y_{3}, W_{1},W_{2},W_{3},W_{4} \}\\
Y_{i} &=  \la \u_{i} \ra, i=1,2,3 \\
W_{1} &= \la \u_{1} + \u_{2} \ra \\ 
W_{2} &= \la \u_{2} + \u_{3} \ra \\
W_{3} &= \la \u_{1} + \u_{2} + \u_{3} \ra \\ 
W_{4} &= \la \u_{1} + \u_{3} \ra . 
\end{align}
where $\u_{1},\u_{2},\u_{3}$ are linearly independent vectors over a
finite field of even characteristic.  Clearly $(\X_1,\Phi_1)$ is even
representable (and hence Ingletonian). In fact, this is the Fano
matroid, $F_7$.

Define the second matroid $(\X_{2}, \Phi_{2})$ with $\Phi_2(\cdot)=\dim\la\cdot\ra$
and 
\begin{align}\label{eq:secondd}
  \X_{2} &\defined \{\seq Z15, \seq V18  \},\\
  Z_{i} &= \la \u_{i} \ra, \: i=1,\ldots, 5 \\
  V_{1} &= \la \u_{1} + \u_{2} + \u_{3} \ra \\
  V_{2} &= \la \u_{3} + \u_{4} + \u_{5} \ra  \\
  V_{3} &= \la \u_{1} + \u_{2}\ra \\
  V_{4} &= \la \u_{1} + \u_{3} \ra  \\
  V_{5} &= \la \u_{2} +  \u_{3}\ra \\
  V_{6} &= \la\u_{3} + \u_{4}\ra  \\
  V_{7} &= \la\u_{3} + \u_{5}\ra \\
  V_{8} &= \la\u_{4} + \u_{5} \ra
\end{align}
where $\{ \u_{1}, \ldots \u_{5} \}$ are linearly independent over a
finite field of odd characteristic. Clearly $(\X_1,\Phi_1)$ is odd
representable (and hence Ingletonian).

It is easy to prove that $(\X_1,\Phi_1)$ and $(\X_2,\Phi_2)$ are both
connected. Furthermore, $(\X_1,\Phi_1)$ satisfies the following
equalities (recalling the notational conventions \eqref{eq:conditionalentropy},~\eqref{eq:conditionalmutualinformation})
\begin{equation}
H(Y_{1},Y_{2},Y_{3}) = \sum_{i=1}^{3}H(Y_{i}) \label{eq:firsta}\\
\end{equation}
\begin{align}    \label{eq:firstb}
    H\left(W_{1}\mid Y_{1}Y_{2}\right) &=0 &
    H\left(W_{2}\mid Y_{2}Y_{3}\right) &=0\\\notag
    H\left(W_{3}\mid Y_{1}W_{2}\right) &=0 &
    H\left(W_{4}\mid W_{1}W_{2}\right) &=0\\\notag
    H\left(Y_{1} \mid Y_{3}W_{4}\right) &=0&
    H\left(Y_{2}\mid W_{3}W_{4}\right) &=0\\\notag
    H\left(Y_{3}\mid W_{1}W_{3}\right) &=0
\end{align}
It is worth pointing out explicitly that $H\left(W_{4} \mid W_{1}W_{2}\right)=0$
because all vectors are defined over a finite field of even
characteristic (hence, $\u_{1}+\u_{3} =\u_{1} + \u_{2} + \u_{2} +
\u_{3}$).  
% Equalities \eqref{eq:firsta} and \eqref{eq:firstb}
% correspond to the ``connection constraint'' in the network depicted in
% Figure \ref{fig:dfz}a. In other words, the so-constructed Ingletonian
% polymatroid $(\X_{1}, \Phi_{1})$ solves the network multicast problem.

Simiarly, $(\X_2,\Phi_2)$  satisfies 
%the following connection constraint for the network depicted in Figure \ref{fig:dfz}b: 
\begin{equation}
  H(\seq Z15) = \sum_{i=1}^{5}H(Z_{i}) \label{eq:seconda}
\end{equation}
  \begin{align}\label{eq:secondb}
    H(V_{1}|Z_{1}Z_{2}Z_{3}) &=0 & H(V_{2}|Z_{3}Z_{4}Z_{5}) &=0 \\\notag
    H(V_{3}|Z_{1}Z_{2}) &=0 & H(V_{4}|Z_{1}Z_{3}) &=0 \\\notag
    H(V_{5}|Z_{2}Z_{3}) &=0 & H(V_{6}|Z_{3}Z_{4}) &=0 \\\notag
    H(V_{7}|Z_{3}Z_{5}) &=0 & H(V_{8}|Z_{4}Z_{5}) &=0 \\\notag
    H(Z_{1}|V_{1}V_{5}) &=0 & H(Z_{2}|V_{1}V_{4}) &=0 \\\notag
    H(Z_{3}|V_{1}V_{3}) &=0 & H(Z_{3}|\seq V38) &=0 \\\notag
    H(Z_{3}|V_{2}V_{8}) &=0 & H(Z_{4}|V_{2}V_{7}) &=0 \\\notag
    H(Z_{5}|V_{2}V_{6}) &=0. 
  \end{align}
  In this case, we emphasis that $H(Z_{3}|\seq V38) =0$ holds because
  the characteristic of the underlying field is odd.
%As before, constraint \eqref{eq:seconda} and \eqref{eq:secondb} correspond to the connection constraint 
%in a network multicast problem. 
%Furthermore, the following triples of pseudo-variables in $(\X_{2},\Phi_{2})$
%\begin{align}\label{eq:secondc}
%\begin{cases}
%(V_{1}, V_{3},Z_ {3}),  (V_{1}, V_{4},Z_{2}),  (V_{1}, V_{5},Z_{1}) \\
%(V_{2}, V_{6},Z_{4}), (V_{2}, V_{7},Z_{5}), (V_{2}, V_{8},Z_{3})
%\end{cases}
%\end{align} 
%are locally symmetric. Once again, for any $(\X_{2} , \bfh)$ that satisfies the local symmetry constraint \eqref{eq:secondc},
%all its psuedo-variables will have the same rank.  

%These set of functional dependencies and independencies can be represented by a network 
%\begin{figure}
%\includegraphics*[scale=0.8]{fig/exist1}
%\caption{DFZ network 2}
%\end{figure}

\begin{thm}\label{thm:existence}
$(\X, \Phi_{1})$ defined by \eqref{eq:firstd} is even representable but not almost odd representable. 
Simiarly, $(\X, \Phi_{2})$ defined by \eqref{eq:secondd} is odd representable but not almost even representable. 
\end{thm}

Before we prove Theorem~\ref{thm:existence}, we will need two lemmas,
which provide some elementary results from linear algebra.
\begin{lemma}\label{lemma:exist1}
  Let $\{ \seq B1n \}$ and $C$ be subspaces of $W$. Then for any
  $\alpha \subseteq \N$, there exists a subspace $A$ such that
 $H\left(A\right) = H\left(C\mid B_{j}, j\in\alpha\right)$ and  $H\left(T_{A}(C) \mid  T_{A}\left(B_{j}\right), j\in\alpha \right)= 0$.
 Furthermore, for a given sequence of subspaces $\{ \seq{\bb{B}^{i}}1n
 , C^{i}\}_{i=1}^{\infty}$ and  $k(i)>0$ such that
 \begin{equation*}
   \lim_{i\to\infty} \frac{1}{k(i)} H\left( C^{i} \mid \bb{B}^{i}_{j}
     , j\in\alpha\right) = 0, 
\end{equation*}
there exists a sequence of subspaces $\{A^{i}\}_{i=1}^{\infty}$ such that for all
$\beta\subseteq \N_{n}$
\begin{align*}
 H\left(A^{i}\right) &= H\left(C^{i}\mid  \bb{B}^{i}_{j} , j\in\alpha\right)\\
 H\left( T_{A^{i}}\left(C^{i}\right) \mid  T_{A^{i}}\left(\bb{B}^{i}_{j}\right) , j\in\alpha\right) &= 0\\
\lim_{i\to\infty} \frac{1}{k\left(i\right)} H\left(T_{A^{i}}\left(\bb{B}^{i}_{j}\right) , j\in\beta\right) &= \lim_{i\to\infty} \frac{1}{k\left(i\right)}H\left(\bb{B}^{i}_{j} , j\in\beta\right) .
\end{align*}
%$\{ \seq{\bb{B}^{i}}1n \}_{i=1}^{\infty} \sim  \{  T_{A^{i}}(\bb{B}^{i}_{1}), \ldots  T_{A^{i}}(\bb{B}^{i}_{n}) \}_{i=1}^{\infty}$. Here, we use the notation  $\{ \seq D1n\} \sim \{ \seq {D^{\p}}1n\}$ to denote that 
%\begin{align}
%\lim_{i\to\infty} H(D_{i} , i\in\alpha) = \lim_{i\to\infty} H(D^{\p}_{i} , i\in\alpha) 
%\end{align}
%for all $\alpha\subseteq \N_{n}$.
\end{lemma}
\begin{proof}
  It is easy to pick a subspace $A$ of $C$ such that $H(A)=H(C\mid B_{j},
  j\in\alpha)$ and that $A$ and $\la B_{j}, j\in\alpha \ra$ together
  span $\la C, B_{j}, j\in\alpha \ra$. Then it is straightforward to
  prove that $H(T_{A}(C) \mid T_{A}(B_{j}) , j\in\alpha )= 0$, which
  proves the first part of the lemma.

  Similarly, for each $i$, there exists a subspace $A^{i}$ such that
  $H(A^{i}) = H( C^{i}\mid \bb{B}^{i}_{j} , j\in\alpha)$, and $H(
  T_{A^{i}}(C^{i}) \mid T_{A^{i}}(\bb{B}^{i}_{j}) , j\in\alpha) = 0$.  By
  Lemma \ref{lemma:property}, for any $\beta$
\begin{align}
H\left(  B^{i}_{\beta}   \right) - H\left(A^{i}\right) \le H\left( T_{A} \left( B^{i}_{j} \right) , j\in\beta  \right) \le  H\left(  B^{i}_{\beta}  \right).
\end{align}
The remaining part of the lemma then follows as $\lim_{i\to\infty}
\frac{1}{k(i)} H(A^{i}) = 0$.
\end{proof}

\begin{lemma}\label{lemma:exist2}
  Let $\{ \seq {\bb{B}}1n \}$ be a collection of subspaces and $\beta
  \subseteq \N$. Then there exists a subspace $A$ such that
\begin{align}
  H\left(T_{A}\left(B_{j}\right) \right) & = H\left(B_{j}\mid  B_{
      \beta\setminus j }\right), \: \forall j \in
  \beta, \label{eq:lemmaexist2a}\\ 
  H\left(T_{A} \left(B_{j}\right), j\in\beta\right) & =
  \sum_{j\in\beta} H\left(T_{A}\left(B_{j}\right)\right)
  , \label{eq:lemmaexist2b}\\ 
  H\left(A\right) &= H\left( {\bb{B}}_{\beta} \right) -
  \sum_{j\in\beta} H\left(B_{j}\mid B_{ \beta\setminus j
    }\right). \label{eq:lemmaexist2c}
\end{align}
Furthermore, for a sequence of subspaces $\{ \seq{\bb{B}^{i}}1n
\}_{i=1}^{\infty}$ and $k(i)>0$ such that
\[
\lim_{i\to\infty} \frac{1}{k\left(i\right)}  \left(H\left(B^{i}_{\beta}\right) - \sum_{j\in\beta}  H\left(B_{j}^{i}\mid B_{ \beta\setminus j }^{i} \right) \right)  = 0.
\]
  there exists a sequence of subspaces $A^{i}$ such that   for all $\alpha \subseteq \N$,
\begin{align*}
 \sum_{j\in\beta}H\left( T_{A^{i}} \left(B^{i}_{j}\right)\right)  &=
 H\left(T_{A^{i}}\left(B^{i}_{j}\right) , j\in\beta\right)\\ 
\lim_{i\to\infty} \frac{1}{k\left(i\right)}
H\left(T_{A^{i}}\left(\bb{B}^{i}_{j}\right) , j\in\alpha\right) &=
\lim_{i\to\infty} \frac{1}{k\left(i\right)}H\left(\bb{B}^{i}_{\alpha}\right) . 
\end{align*}
\end{lemma}
\begin{proof}
  Define $A$ as the minimal subspace containing $B_{i} \cap \la B_{j},
  j \in \beta\setminus i \ra $ for $i\in\beta$.  Then it is
  straightforward to prove that for all $i$, $A$ is a subspace of $\la
  B_{j}, j\in \beta\setminus i \ra$ and hence $H(A | B_{j}, j\in
  \beta\setminus i ) = 0$. Similarly, for all $i$,
 $B_{i} \cap A = B_{i} \cap \la  B_{j}, j\in \beta\setminus i \ra$ and hence
$I(B_{i} ; A) = I(B_{i} ;  B_{j}, j\in \beta\setminus i )$ and 
\begin{equation*}
H\left(\seq B1n \mid  A\right) = \sum_{i=1}^{n} H\left(B_{i}\mid A\right) = H\left(B_{i}\mid B_{ \beta\setminus i }\right)
\end{equation*}
Consequently, \eqref{eq:lemmaexist2a}-\eqref{eq:lemmaexist2c}
holds. The remaining part of the lemma be proved similarly as in Lemma
\ref{lemma:exist1}.
 \end{proof}

%----------------------------------------------------------------------
% DFZ rank ratio results
%
 The final ingredients that we require are the following results from
 \cite{Dougherty.Freiling.ea05insufficiency}. Although these results
 were originally stated in terms of linear network coding capacity, we
 can restate them purely in terms of rank inequalities as follows.
\begin{thm}[Dougherty, Freiling, Zeger]\label{thm:dfz}
  Suppose that $\X_{1} = \{ Y_{1},Y_{2},Y_{3}, \seq W14\}$ is a
  collection of vector subspaces %(induced by global encoding kernels)
  over a finite field of odd characteristic. If the resulting
  polymatroid satisfies 
%(network topology and connection  requirement) 
 \eqref{eq:firsta} and \eqref{eq:firstb},
  then~\cite[Theorem IV.3]{Dougherty.Freiling.ea05insufficiency}
\[
\frac{\min_{i=1,2,3} H (Y_{i}) }{\max_{j=1,2,3,4} H(W_{j})} 
\le 
\frac{4}{5}.
\] 
Similarly, suppose that $\X_{2} = \{ \seq Z15, \seq V18\}$ is a
collection of vector subspaces over a finite field of even
characteristic. If $\X_{2}$ satisfies \eqref{eq:seconda} and
\eqref{eq:secondb}, then~\cite[Theorem IV.4]{Dougherty.Freiling.ea05insufficiency}
\[
\frac{\min_{i=1,\ldots,5} H (Z_{i}) }{\max_{j=1,\ldots,8} H(V_{j})} 
\le 
\frac{10}{11}.
\] 
\end{thm}

 \begin{proof}[Proof of Theorem~\ref{thm:existence}]
   Suppose to the contrary that $(\X, \Phi_{2})$ is almost even
   representable. Then by definition there exists a sequence of even
   representable polymatroids $\{(\X, \bfg_{i} )\}_{i=1}^{\infty}$ and
   positive constants $d_{i}$ such that $\lim_{i\to\infty} d_{i}
   \bfg_{i} = \Phi_{2}$.  While $(\X, \Phi_{2})$ satisfies
   \eqref{eq:seconda} and \eqref{eq:secondb}, these constraints may
   not be satisfied by $(\X, \bfg_{i})$ in general. However, we can
   use Lemmas \ref{lemma:exist1} and \ref{lemma:exist2} to construct
   from $\{(\X, \bfg_{i} )\}_{i=1}^{\infty}$ another sequence of even
   representable polymatroids $\{(\X, \bfg_{i}^{\p} )\}_{i=1}^{\infty}$
   such that $ \bfg_{i}^{\p}$ satisfies \eqref{eq:seconda} and
   \eqref{eq:secondb}, and $\lim_{i\to\infty} {d_{i}} \bfg_{i}^{\p} =
   \lim_{i\to\infty}{d_{i}} \bfg_{i} = \Phi_{2}$.  As such,
   \begin{align}\label{eq:limit1}
     \lim_{i\to\infty }\frac{\min_{k=1, \ldots , 5} {\bfg_{i}^{\p}}(Z_{k})
     }{\max_{k=1,\ldots,8} {\bfg_{i}^{\p}}(V_{k}) } = \frac{\min_{i=1,
         \ldots , 5} {\Phi_{2}}(Z_{k}) }{\max_{k=1,\ldots,8}
       {\Phi_{2}}(V_{k}) } \nequal{(a)} 1,
   \end{align}
   where (a) follows from connectivity of $(\X, \Phi_{2})$. The
   existence of such a sequence $\{(\X, d_{i}\bfg_{i}^{\p}
   )\}_{i=1}^{\infty}$ contradicts Theorem \ref{thm:dfz} which proved
   that the limit \eqref{eq:limit1} is bounded above by $10/11$.  Thus
   $(\X, \Phi_{2})$ cannot be almost even representable.  Using the
   same argument, we can also prove that $(\X, \Phi_{1})$ is not
   almost odd representable. 
 \end{proof}

%%%%%%%%%%%%%%%%%%%%%%%%%%%%%%%%%%%%%%%%%%%

 \section{Insufficiency of All Known Rank Inequalities}\label{sec:insufffive}

 In Section \ref{sec:newpoly}, we constructed $(\X,\Phi
 )\in\barcon(\Upsilon[\X])$ such that its $\epsilon$-pertubation
 $(\X,\Phi^{\epsilon} )\not\in\barcon(\Upsilon[\X])$.  Theorem
 \ref{thm:extension} proved the existence of new subspace rank
 inequalities that are not implied by Ingleton's inequalities. This
 was achieved by showing that $(\X,\Phi^{\epsilon})$ is
 Ingletonian. In the following, we will give another proof for Theorem
 \ref{thm:extension}. This alternative proof demonstrates the kind of
 difficulties one may face when characterizing representable
 polymatroids. Finally, we will generalize our main result to show
 that the newly discovered DFZ inequalities are also insufficient.

\begin{proof}[Alternative proof of Theorem  \ref{thm:extension}]
  In \cite{Zhang.Yeung98characterization} and
  \cite{Hammer2000Inequalities}, all the extreme vectors
  $\barcon(\Upsilon[\X])$ are identified for $|\X| = 4$. It can be
  easily verified that all of the associated rank functions are
  ternary representable. Hence, every vector in
  $\barcon(\Upsilon[\X])$ is almost representable. By Theorem
  \ref{thm:perturbrepresentable}, its $\epsilon$-perturbed counterpart
  is almost representable.

  Now, consider any Ingletonian polymatroid $(\X, \bfh)$ and its
  $\epsilon$-perturbation $(\X, \bfg)$ defined as in
  \eqref{eqn:extension}.  For any subsets $\V_{1},\ldots, \V_{4}\subseteq\X$,
  $\bfh$ induces a polymatroid $(\N_4,\bfh')$ via $\bfh^{\p}(\alpha)
  \defined \bfh(\V_{i},i\in\alpha)$.  Clearly, $\bfh^{\p}$ is also
  Ingletonian, since for any
  subsets $\alpha_1,\dots,\alpha_4\subseteq\N_4$, $J_{\bfh^{\p}} (
  \alpha_{1}, \ldots, \alpha_{4} ) = J_{\bfh} ( \V_{\alpha_{1}},
  \ldots, \V_{\alpha_{4}} ) $. Hence, $\bfh^{\p}$ is also almost
  representable. By Theorem \ref{thm:perturbrepresentable}, its
  perturbation is also almost representable.

  Now, if $\bfh (\X)-\epsilon \ge \bfh^{\p}(\N_4)$, then clearly
  \begin{align}
    \bfg (\V_{i} , i\in\alpha)  = \bfh (\V_{i} , i\in\alpha) 
  \end{align}
  for all $\alpha$. Then $ J_{\bfg} ( \V_{\alpha_{1}}, \ldots,
  \V_{\alpha_{4}} ) = J_{\bfh} ( \V_{\alpha_{1}}, \ldots,
  \V_{\alpha_{4}} ) \ge 0 $.  

  On the other hand, suppose that $\bfh (\X)-\epsilon \le
  \bfh^{\p}(\N_4)$. Then
\begin{align}
\bfg (\V_{i} , i\in\alpha)  & =   \min ( \bfh (\V_{i} , i\in\alpha),  \bfh (\X)-\epsilon) \\
& =   \min ( \bfh^{\p} (\alpha),  \bfh (\X)-\epsilon) \\
& =   \min ( \bfh^{\p} (\alpha),  \bfh^{\p}(\N_4) + \bfh(\X) - \bfh^{\p}(\N_4) -\epsilon) \\
& =   \min ( \bfh^{\p} (\alpha),  \bfh^{\p}(\N_4)  -\nu  ) 
\end{align} 
where $\nu \defined \bfh^{\p}(\N_4) +\epsilon - \bfh(\X) $.
Let 
\[
\bfg^{\p}(\alpha) \defined \min ( \bfh^{\p} (\alpha),  \bfh^{\p}(\N_4)  -\nu  ).
\]
Then $\bfg^{\p}$ is almost representable and hence $J_{\bfg} ( \V_{\alpha_{1}}, \ldots, \V_{\alpha_{4}} ) =  J_{\bfg^{\p}} ( \alpha_{1}, \ldots, \alpha_{4} ) \ge 0 $.
\end{proof}

\begin{thm}[Generalization]\label{thm:generalization}
  Suppose $|\Y| = n$ and that all vectors $ \bfg \in
  \barcon(\Upsilon[\Y])$ are almost representable.  Consider any valid
  subspace rank inequality of the form
\begin{align}\label{eq:generalization}
\sum_{\alpha \subseteq \N_{n}} c_{\alpha} H(\V_{i}, i\in\alpha) \ge 0
\end{align}
where $\V_{i} \subseteq \Y$.  Then for any cc-representable
polymatroid $(\X, \bfh)$ (i.e., $\bfh \in \barcon(\Upsilon[\X])$), its
perturbed counterpart will satisfy the inequality
\eqref{eq:generalization}.  Consequently $(\X,\Phi^{\epsilon})$ in
Section \ref{sec:constructed} will satisfy \eqref{eq:generalization}
and hence any inequalities involving no more than $n$ subsets are
insufficient to characterize $ \barcon(\Upsilon[\X])$ in general.
\end{thm}
\begin{proof}
The proof is similar to the alternative proof for Theorem \ref{thm:main}.
\end{proof}

The newly discovered inequalities~\cite{Dougherty2009Non-Shannon} were shown to be sufficient
to characterize $ \barcon(\Upsilon[\X])$ when $|\X| \le 5$.  In fact,
it was further proved that all the extreme vectors of the cone $
\barcon(\Upsilon[\X])$ are $q$-representable, for sufficiently large
$q$. As a result, every vector in $ \barcon(\Upsilon[\X])$ will be
almost representable.  By Theorem \ref{thm:generalization}, these
newly discovered inequalities are insufficient to characterize $
\barcon(\Upsilon[\X])$ in general.
 
%\begin{thm}
%Any subspace rank inequalities involving less than or equal to 5 variables are not sufficient.
%\end{thm} 
%%%%%%%%%%%%%%%%%%%%%%%%%%

\section{Conclusion}\label{sec:conclusion}
A complete characterization of representable polymatroids has been
open for years. This problem is fundamental in nature and is
intimately related to the information thoeretic problem of the
characterization of transmission throughput in networks with linear
network coding. Until quite recently it was not know whether
Ingleton's inequalities are sufficient to characterize all
representable polymatroids.  In this paper, we have constructed an
Ingletonian polymatroid that satisfies all known (Ingleton and
Dougherty-Freiling-Zeger) subspace rank inequalities. As a result,
there are inequalities remaining to be discovered. While our approach
does not suggest how to construct these new inequalities, it at least
demonstrates some of the difficulties of the problem.

%In this paper, we prove the existence of an Ingletonian polymatroid via two network multicast problems. Such a polymatroid is then proved lying outside the convex hull of all cc-representable polymatroids. 
%This solves the long-standing open problem by showing that Ingleton inequalities are not sufficient to characterize all representable polymatroids.
% 
\section*{Acknowledgment}

This work was supported by the Australian Government under ARC grant
DP0880223. 

\bibliographystyle{IEEEtran}

\end{document}